\documentclass[conference]{IEEEtran}
\IEEEoverridecommandlockouts
\usepackage{cite}
\usepackage{amsmath,amssymb,amsfonts}
\usepackage{algorithmic}
\usepackage{graphicx}
\usepackage{textcomp}
\usepackage[dvipsnames]{xcolor}
\usepackage{todonotes}
\usepackage{hyperref}
\usepackage[nolist,nohyperlinks]{acronym}
\usepackage{tikz}
\usepackage{pbox}
\usepackage{listings}
\usepackage{lstautogobble}
\usepackage{booktabs}
\usepackage{pgfplots}
\usepackage{pgfplotstable}
\usepackage{float}
\usepackage{inconsolata}
\usepackage[T1]{fontenc}
\usepackage[2023, cc-by, pages={7}{12}, url=https://osda.ws/r/2QkTi, auto=false]{osda}
\usetikzlibrary{shapes,arrows, positioning, decorations.markings}

\hypersetup{colorlinks=true, linkcolor=Black, citecolor=Black, urlcolor=black}

\newacro{DP}{Dynamic Programming}
\newacro{FPGA}{Field-Programmable Gate Array}
\newacro{ASIC}{Application-Specific Integrated Circuit}
\newacro{HEV}{Hybrid Electric Vehicle}
\newacro{HDL}{Hardware Description Language}
\newacro{HLS}{High Level Synthesis}
\newacro{GPU}{Graphics Processing Unit}
\newacro{V2X}{Vehicle to Everything}
\newacro{SOC}{State of Charge}
\newacro{VLIW}{Very Large Instruction Word}
\newacro{FWL}{Fractional Word Length}
\newacro{ALM}{Adaptive Logic Module}
\newacro{IWL}{Integer Word Length}
\newacro{FWL}{Fractional Word Length}
\newacro{HDL}{Hardware Description Language}
\newacro{HLS}{High-Level Synthesis}
\newacro{HIR}{High-Level Intermediate Representation}
\newacro{MIR}{Medium-Level Intermediate Representation}
\newacro{DSL}{Domain Specific Language}
\newacro{LSP}{Language Server Protocol}
\newacro{DUT}{Design Under Test}
\newacro{AST}{Abstract Syntax Tree}
\newacro{HIR}{High-Level Intermediate Representation}
\newacro{MIR}{Mid-Level Intermediate Representation}

\usepackage{xifthen}

\ifthenelse{\isundefined{\darktheme}}%
{%
    \definecolor{keywordColor}{rgb}{0.08, 0.22, 0.51}
    \definecolor{commentGreen}{rgb}{0,0.4,0}
    \definecolor{stringColor}{rgb}{0.58,0,0.82}
    \definecolor{bgColor}{rgb}{1,1,1}
}%
{%
    \definecolor{keywordColor}{rgb}{0, 0.68, 0.68}
    \definecolor{commentGreen}{rgb}{0.32, 0.6, 0.38}
    \definecolor{stringColor}{rgb}{0.58,0,0.82}
    \definecolor{bgColor}{rgb}{0,0,0}
}

\ifdefined\codesize
\else
\newcommand{\codesize}{\small}
\fi

\lstdefinelanguage{pseudorust}{
    morekeywords=
        { if
        , else
        , for
        , in
        , let
        , fn
        , type
        , each
        , return
        , is
        , not
        , and
        },
    sensitive=true, 
    morecomment=[l]{//}, 
    morecomment=[s]{/*}{*/}, 
    morestring=[b]", 
}

\ifthenelse{\isundefined{\darktheme}}%
{%
    \definecolor{spadeKeyword}{rgb}{0.16, 0.36, 0.60}
    \definecolor{spadeType}{rgb}{0.84, 0.0, 0.37}
    \definecolor{spadeConditional}{rgb}{0.46, 0.28, 0.58}
    \definecolor{spadeBuiltin}{rgb}{0.84, 0.37, 0.0}
    \definecolor{spadeOperator}{rgb}{0, 0, 0}
    \definecolor{stringColor}{rgb}{0.37, 0.0, 1.0}
    \definecolor{spadeTypeOperator}{rgb}{0, 0.37, 0.52}
    \definecolor{spadeBoolLiteral}{rgb}{0.03, 0.54, 0.03}
    \definecolor{bgColor}{rgb}{1,1,1}

    \definecolor{errorRed}{rgb}{0.81,0.18,0.14}
    \definecolor{errorBlue}{rgb}{0,0.24,0.55}
    \definecolor{errorOrange}{rgb}{0.72,0.62,0.03}
}%
{%
    \definecolor{spadeKeyword}{rgb}{0, 0.78, 0.78}
    \definecolor{spadeType}{rgb}{0.67, 0.81, 0.17}
    \definecolor{spadeConditional}{rgb}{0.84, 0.57, 0.84}
    \definecolor{spadeBuiltin}{rgb}{1.0, 0.37, 0.68}
    \definecolor{spadeOperator}{rgb}{0.88, 0.65, 0.10}
    \definecolor{spadeTypeOperator}{spadeOperator}
    \definecolor{stringColor}{rgb}{0.55, 0.71, 0.33}
    \definecolor{spadeBoolLiteral}{rgb}{0.88, 0.65, 0.10}
    \definecolor{bgColor}{rgb}{0,0,0}
    \definecolor{errorRed}{red!40}
    \definecolor{errorBlue}{blue!40}
    \definecolor{errorOrange}{orange!40}
}

\lstdefinestyle{highlight}{
    keywordstyle=[1]\color{spadeKeyword},
    keywordstyle=[2]\color{spadeType},
    keywordstyle=[3]\color{spadeConditional},
    keywordstyle=[4]\color{spadeBuiltin},
    keywordstyle=[5]\color{spadeOperator},
    keywordstyle=[6]\color{spadeBoolLiteral},
    keywordstyle=[7]\color{spadeTypeOperator},
    commentstyle=\color{commentGreen},
    stringstyle=\color{stringColor},
}
\lstdefinestyle{base}{
    language=Spade,
    basicstyle=\color{black!40}\codesize\ttfamily,
    keywordstyle=[1]\color{black!40},
    keywordstyle=[2]\color{black!40},
    keywordstyle=[3]\color{black!40},
    keywordstyle=[4]\color{black!40},
    keywordstyle=[5]\color{black!40},
    keywordstyle=[6]\color{black!40},
    keywordstyle=[7]\color{black!40},
    commentstyle=\color{black!40},
    moredelim={**[is][\only<0>{\color{white}\lstset{style=highlight}}]{@0}{0@}},
    moredelim={**[is][\only<1>{\color{white}\lstset{style=highlight}}]{@1}{1@}},
    moredelim={**[is][{\only<2>{\color{white}\lstset{style=highlight}}}]{@2}{2@}},
    moredelim={**[is][{\only<3>{\color{white}\lstset{style=highlight}}}]{@3}{3@}},
    moredelim={**[is][{\only<4>{\color{white}\lstset{style=highlight}}}]{@4}{4@}},
    moredelim={**[is][{\only<5>{\color{white}\lstset{style=highlight}}}]{@5}{5@}},
    moredelim={**[is][{\only<6>{\color{white}\lstset{style=highlight}}}]{@6}{6@}},
    moredelim={**[is][{\only<7>{\color{white}\lstset{style=highlight}}}]{@7}{7@}},
    moredelim={**[is][{\only<8>{\color{white}\lstset{style=highlight}}}]{@8}{8@}},
    moredelim={**[is][{\only<9>{\color{white}\lstset{style=highlight}}}]{@9}{9@}},
    moredelim={**[is][{\only<4->{\color{white}\lstset{style=highlight}}}]{@on4}{4on@}},
    moredelim={**[is][{\only<5->{\color{white}\lstset{style=highlight}}}]{@on5}{5on@}},
}

\lstdefinestyle{error}{
    moredelim={**[is][\color{errorRed}]{@r}{r@}},
    moredelim={**[is][\bfseries\color{errorBlue}]{@b}{b@}},
    moredelim={**[is][\color{errorOrange}]{@y}{y@}},
}

\lstdefinestyle{spadeinline}{
    language=Spade,
    basicstyle=\color{black}\ttfamily,
    keywordstyle=[1]\color{black},
    keywordstyle=[2]\color{black},
    keywordstyle=[3]\color{black},
    keywordstyle=[4]\color{black},
    keywordstyle=[5]\color{black},
    keywordstyle=[6]\color{black},
    keywordstyle=[7]\color{black},
    commentstyle=\color{black!40},
}

\lstdefinelanguage{spade}{
    basicstyle=\codesize\ttfamily\mdseries,
    keywords=[1]{
        entity, fn, reg, let, reset, inst, enum, decl, struct, use, as, mod, stage, pipeline, assert, port, set
    },
    keywordstyle=[1]\color{spadeKeyword}\ttfamily\bfseries,
    keywords=[2]{int,bool,Option,clock},
    keywordstyle=[2]\color{spadeType},
    keywords=[3]{match, if, else},
    keywordstyle=[3]\color{spadeConditional}\ttfamily\bfseries,
    keywords=[4]{Some, None},
    keywordstyle=[4]\color{spadeBuiltin}\ttfamily\bfseries,
    otherkeywords={=,-,<,>,+,|},
    keywords=[5]{=,-,<,>,+,|},
    keywordstyle=[5]\color{spadeOperator}\ttfamily\bfseries,
    keywords=[6]{true,false},
    keywordstyle=[6]\color{spadeBoolLiteral},
    otherkeywords={&},
    keywords=[7]{&,mut},
    keywordstyle=[7]\color{spadeTypeOperator}\ttfamily\bfseries,
    sensitive=true, 
    morecomment=[l]{//}, 
    morecomment=[s]{/*}{*/}, 
    commentstyle=\color{commentGreen},
    morestring=[b]", 
    escapechar=\%,
}

\lstdefinelanguage{python}{
    otherkeywords={&, @,=},
    keywords=[1]{
        async, def, await
    },
    keywordstyle=[1]\color{spadeKeyword}\ttfamily\bfseries,
    keywords=[2]{int, bool},
    keywordstyle=[2]\color{spadeType},
    keywords=[2]{int, bool},
    keywords=[3]{match, if, else},
    keywordstyle=[3]\color{spadeConditional},
    keywords=[4]{Some, None, true, false}
    keywordstyle=[4]\color{spadeBuiltin},
    keywords=[5]{@,=},
    keywordstyle=[5]\color{spadeOperator},
    sensitive=true, 
    morecomment=[l]{\#}, 
    morecomment=[s]{/*}{*/}, 
    commentstyle=\color{commentGreen},
    morestring=[b]{"}, 
    stringstyle=\color{stringColor},
    escapechar=\%,
}

\newfloat{lstfloat}{tpb}{lop}
\floatname{lstfloat}{Listing}

\newcommand{\spadeinline}[1]{\lstinline[language=spade, style=spadeinline, breaklines=true,breakatwhitespace]{#1}}

\usetikzlibrary{shapes,arrows, positioning, decorations.markings}

\tikzstyle{decision} = [diamond, draw, fill=blue!20, 
    text width=4.5em, text badly centered, node distance=3cm, inner sep=0pt]
\tikzstyle{process} = [block, ]
\tikzstyle{computation} = [circle, draw]
\tikzstyle{pipeline} = [block, draw]
\tikzstyle{artefact} = [trapezium, draw, text centered, trapezium left angle=120, trapezium right angle=60, ]

\tikzstyle{block} = [rectangle, draw, text centered, ]
\tikzstyle{ring} = [circle, draw]

\tikzstyle{textOnly} = [rectangle, text centered, ]
\tikzstyle{arith} = [circle, inner sep=0.05cm, draw]
\tikzstyle{pipelineReg} = [block, draw, text width=2.5em]
\tikzstyle{invisible} = [textOnly, inner sep=0, minimum size=0]

\tikzstyle{line} = [draw]
\tikzstyle{arrow} = [draw, -latex']
\tikzstyle{backarrow} = [draw, latex'-]

\definecolor{RYB1}{RGB}{31,  94,  255}
\definecolor{RYB2}{RGB}{255, 31,  94}
\definecolor{RYB3}{RGB}{49,  173, 0}
\definecolor{RYB4}{RGB}{251, 128, 114}
\definecolor{RYB5}{RGB}{128, 177, 211}
\definecolor{RYB6}{RGB}{253, 180, 98}
\definecolor{RYB7}{RGB}{179, 222, 105}

\tikzstyle{LLBasicBlock} = [rectangle, draw, align=left]

\tikzset{%
  dots/.style args={#1per #2}{%
    line cap=round,
    dash pattern=on 0 off #2/#1
  }
}

\def\BibTeX{{\rm B\kern-.05em{\sc i\kern-.025em b}\kern-.08em
    T\kern-.1667em\lower.7ex\hbox{E}\kern-.125emX}}

\renewcommand{\codesize}{\footnotesize}

\title{Spade: An Expression-Based HDL With Pipelines}

\author{\IEEEauthorblockN{Frans Skarman and Oscar Gustafsson}
\IEEEauthorblockA{Department of Electrical Engineering, Linköping University\\
SE-581 83 Linköping, Sweden\\
Email: \{frans.skarman, oscar.gustafsson\}@liu.se}
}

\begin{document}
\bstctlcite{BSTcontrol}

\maketitle

\begin{abstract}
    Spade is a new open source hardware description language (HDL) designed to
    increase developer productivity without sacrificing the low-level control
    offered by HDLs. It is a standalone language which takes inspiration from modern software
    languages, and adds useful abstractions for common hardware constructs.
    It also comes with a convenient set of tooling, such as a helpful compiler,
    a build system with dependency management, tools for debugging, and editor
    integration.
\end{abstract}


\section{Introduction}

Developing digital hardware is traditionally done using Verilog or VHDL, both languages originating in the 1980s.
While they have evolved since their inception, they are still lacking many subsequent advancements in language design.

Spade\footnote{\url{https://gitlab.com/spade-lang/spade/}} is a new open source \ac{HDL} which aims to reduce the development effort of digital hardware by taking inspiration from software languages, and adding language-level
mechanisms for common hardware structures. This is while still retaining the low-level
control provided by \acp{HDL}.

Being inspired by Rust and functional programming languages, Spade
is expression-based and has a rich type system. It supports
product-types like structs and tuples, and sum-types in the form of enums.
The language also supports linear type-checking which can be used to ensure
that hardware resources such as memory ports are used exactly once.
\osdafootnote{}%

Spade has built in constructs and abstractions for common hardware structures such
as pipelines, memories, and registers.
Pipelining allows the user to specify a computation to be performed, with explicit
statements for separating stages of the pipeline, but without the need of separate variables for each pipeline stage.
Retiming such a pipeline does not require changing any variables, only moving
the staging statement.
Additionally, the delays of pipelines are explicit in the language and checked
by the compiler to ensure that changes to a pipeline do not affect the computation results.
In order to more accurately reflect the hardware being described, all logic
is combinatorial by default with the only sequential elements being registers
and memories which are instantiated explicitly.

The rest of the paper is structured as follows. Related works are discussed in Section~\ref{sec:related_work}, including highlighting where Spade differs.
The basic semantics are introduced in Section~\ref{sec:basic_semantics}, while in
Section~\ref{sec:ports} the use of linear types to model input and output ports is described. The software provided for Spade is discussed in Section~\ref{sec:software}, with concluding remarks in Section~\ref{sec:conclusions}.

\section{Related Work}\label{sec:related_work}

In recent years, several new \acp{HDL} have been developed.
A common approach is to embed the language as a \ac{DSL} inside
a conventional software programming language such as Scala, Python, or Ruby.
There, the \ac{HDL} consists of a library of hardware constructs which the user instantiates
in the host language in order to describe their hardware.
This allows the user to take advantage of the power of the host language in
their hardware description, for example, by using software control flow
structures to generate parameterized hardware and using object-oriented or functional
programming approaches in the hardware description. Additionally, this approach
exposes all the tooling available for the host language to the hardware designer, such as
dependency managers, build tools, and IDEs. Examples of this include
Chisel~\cite{bachrach:2012:chisel}, SpinalHDL~\cite{SpinalHDLGithub} and DFiant~\cite{Port2017}
embedded in Scala, Amaranth~\cite{AmaranthGithub} embedded in Python,
ROHD~\cite{Korbel2022} embedded in Dart, and RubyRTL~\cite{LeLann20} embedded
in Ruby.

Some languages take subsets of conventional programming languages and compile
them to hardware.
An example of this is Clash~\cite{Baaij2015} which compiles a large subset of
Haskell to hardware by taking advantage of the natural mapping between a pure
functional language and hardware.
Another example is PipelineC~\cite{Kemmerer2022} which is a
C-like \ac{HDL} with automatic pipelining.
While not regular C, it is close enough to it to allow many PipelineC
programs to be compiled and ``simulated'' by a standard C-compiler.

TL-Verilog~\cite{Hoover2017} is a SystemVerilog extension which supports
timing abstract modeling, where behavior and timing are separated. It provides
language features for common hardware constructs such as pipelines and FIFOs.

There are also several languages which are completely independent of existing languages.
One such example is Bluespec~\cite{Nikhila2004} in which hardware is described
by guarded atomic actions.
Another example is Silice~\cite{SiliceNov2022} which contains abstractions for
common hardware constructs without losing control over the generated hardware.
In addition, it provides a higher-level description style where the design is expressed
as sequences of operations with software-like control flow.

Finally, a common alternative for describing hardware is \ac{HLS}, in which
higher-level languages, typically designed for software, are compiled to
hardware.
This design methodology is quite different from the \acp{HDL}
described previously. Specifically, \ac{HLS} tools generally provide limited
control over the hardware being generated.
In particular, the exact timing of any circuit is generally abstracted away,
and synchronization between \ac{HLS} generated modules is done at runtime via
synchronization signals.

Now, a natural question to ask is whether there
is a need for another \ac{HDL}, and what Spade offers that the existing work
does not?
First, it is not an embedded \ac{DSL} in another language, which separates
it from the likes of Chisel and Amaranth. This gives more freedom in the design
of the language, as it is not restricted to the expressiveness of a \ac{DSL}.
For example, most embedded HDLs are forced to invent new
``keywords'' in order to not clash with the host language, a typical example being Chisel using
\texttt{when}, and \texttt{otherwise} instead of the more familiar \texttt{if},
and \texttt{else}.
This problem runs deeper than surface level keywords, however.
As an example, Scala has pattern matching, but in Chisel and SpinalHDL, that
feature can only be used at ``build time'' on Scala values, not in hardware on
``runtime'' values.
Finally, a potential user is not required to learn both the host language and
\ac{DSL} when initially picking up the language.

Unlike Clash and PipelineC, Spade is not restricted to conforming
to the execution model of the host language which means that the language can be
designed directly for hardware description without any restrictions.

Spade describes the behavior of circuits in a cycle-to-cycle manner, which gives more control than
Bluespec, where the fundamental abstraction is atomic rules, and
PipelineC where the compiler automatically performs pipelining.

Finally, Spade is similar in spirit to TL-Verilog.
However, by designing a new language instead of building on SystemVerilog,
there is more freedom to explore new language constructs.

\section{Basic Semantics}\label{sec:basic_semantics}

In order to describe the programming model and semantics of Spade, an example is used.
Listing~\ref{lst:blink} shows Spade code which blinks an
LED at a configurable interval.
Lines~1--2 define an \spadeinline{entity} called blink which takes a clock
signal of type \spadeinline{clock}, a reset signal of type \spadeinline{bool}, and a 20-bit integer, \spadeinline{int<20>}, specifying the maximum value for the counter.
The return value is a single \spadeinline{bool} value which drives the led.
The \spadeinline{entity} keyword specifies that this \textit{unit} can contain
sequential logic.
One can also write a function, \spadeinline{fn}, which only allows
combinatorial logic, and \spadeinline{pipeline} which defines a pipeline.
The details of pipelines are discussed later.
When instantiating a unit, the
syntax differs between functions, entities and pipelines, giving a reader of the code
an indication of what can happen behind an instantiation.

\begin{lstfloat}
\begin{lstlisting}[ language=spade
        , numbers=left
        , numberstyle=\color{gray}
        , xleftmargin=2em
        , caption=Spade code which blinks an LED\label{lst:blink}
        ]
entity blink(clk: clock, rst: bool, max: int<20>)
    -> bool
{
    reg(clk) counter reset (rst: 0) =
        if counter == max {
            0
        } else {
            trunc(counter + 1)
        };
    counter > (max >> 1)
}
\end{lstlisting}
    \vspace{-0.7cm}
\end{lstfloat}

Line~4 defines a register called \spadeinline{counter} which is clocked by the
\spadeinline{clk} signal and reset to 0 by the \spadeinline{rst} signal.
Registers in Spade are explicit constructs rather than being inferred from the
use of something like \spadeinline{rising\_edge}.
Registers are, together with writes to memories, the only sequential constructs
in the language, everything else describes combinatorial circuits.

In Spade, the behavior of a circuit is described in a cycle-to-cycle manner.
The new value of a register in the design is given as an expression of the
register values in the current clock cycle.
All variables in Spade are immutable: they can only be assigned once.
Single assignment is possible because Spade is an expression-oriented language, meaning that
most statements are expressions and produce values\cite{Klabnik2018}.
For example, the \spadeinline{counter} register is set to
the value of the expression following the equals sign on line~4, in this case
an if-expression spanning lines~5--10.
If the counter has reached the max value, the value \textit{returned} by the if expression
is 0, otherwise it is the next value of the counter.
Compared to traditional imperative \acp{HDL}, the use of immutable variables and
expression-based control flow is closer to the resulting hardware where
variables correspond to wires and ``control flow'' is implemented with
multiplexers.
The language also does not have an explicit return keyword, the last expression
in a unit body is the output of the entity, here on line~11.

The call to the \spadeinline{trunc} function on line 9 truncates the value of \spadeinline{(counter + 1)} from 21 bits down to 20 bits.
This is needed as Spade prevents accidental overflow by using the largest possible
values for most arithmetic operations.
More details on the type system is given in Section~\ref{sub:types}

Spade has similar scoping rules to most modern software languages: variables are only visible
below their definition which makes it more difficult to accidentally create
combinatorial loops.
Additionally, these scoping rules make it easier for a developer to find the definition of a variable.
The definition will always be above where it is used, and generally be grouped
with its assignment.
Sometimes it is still useful to create a loop of dependencies between
variables, however, this is done explicitly using the \spadeinline{decl} keyword.

\subsection{Pipelines}

Pipelining is an important construct in most hardware designs, it allows designs
to maintain a high clock frequency and throughput at the cost of latency.
However, despite their importance, most \acp{HDL} require the user to manually
build their pipelines, a process that is tedious and error-prone as one must make
sure that computations are performed on values corresponding to the correct time step.
In some cases, designers use patterns on their variable names, and ad-hoc static checking
tools to verify that the pipelining is correct~\cite{LuuWritingSafeVerilog}.

\begin{lstfloat}[t]
    \begin{lstlisting} [ language=spade
        , autogobble=true
        , caption={Example of the pipelining construct in Spade.}
        , label=lst:pipeline
        , numbers=left
        , numberstyle=\color{gray}
        , xleftmargin=2em
        ]
        pipeline(4) X(clk: clock, a: int<32>, b: int<32>)
            -> int<33>
        {
                'initial
                let x = inst(3) subpipe(clk, a);
                let p = a * b;
            reg * 3;
                let s = x + f(a, p);
            reg;
                s + stage(initial).a
        }

        pipeline(3) subpipe(...) -> int<32> {...}
    \end{lstlisting}
\end{lstfloat}

\begin{figure}
    \centering
    \includegraphics[]{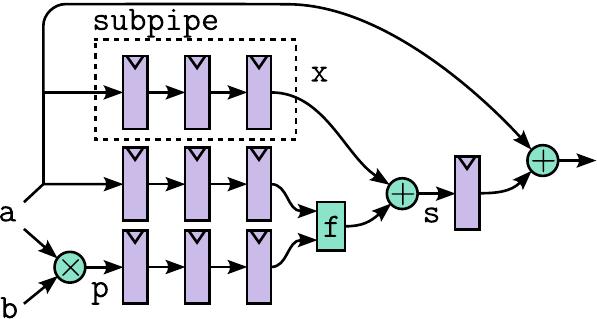}
    \caption{Hardware described in pipeline \spadeinline{X} by the code in Listing~\ref{lst:pipeline}.}\label{fig:pipeline}
    \vspace{-0.5cm}
\end{figure}

Spade on the other hand has language-level support for describing pipelines
where the user describes which computations are performed in each stage,
and the compiler manages the insertion of registers between stages.
Listing~\ref{lst:pipeline} shows an example of the pipelining construct in use to describe the
hardware in Fig.~\ref{fig:pipeline}.
Like the \spadeinline{entity} discussed previously, the first two lines describe the
external interface of the pipeline: name, inputs, and outputs.
It also specifies depth, in this case 4, which is the number of registers in the
pipeline and therefore the delay between input and output.
While the compiler can infer the depth, it has to be specified manually
because it is part of the public API.
This allows a user to see the interface of the pipeline without reading the body.

Line 4 names the current pipeline stage \spadeinline{initial}, allowing references
to it later in the code.
On line~5, another pipeline called \spadeinline{subpipe} is instantiated using the \spadeinline{inst} keyword.
When instantiating pipelines, the depth of the pipeline has to be specified, and the compiler
checks this against the depth specified in the head of the instantiated pipeline.
If the depth of the instantiated pipeline is changed later, the compiler
gives an error, forcing the developer to consider whether that change
has an effect in the instantiating code.
The compiler also checks the availability of variables from sub-pipelines, giving
an error if the result is used before it is ready. As an example, if \spadeinline{x} is
used in the stage right after its definition, it will emit the error message shown in Listing~\ref{lst:pipeline_error}.

Line 6 defines a new variable called \spadeinline{p}, which is the product of \spadeinline{a} and \spadeinline{b}, and, from type inference, an \spadeinline{int<64>}.
Following that is a \spadeinline{reg}-statement, which behaves differently from the \spadeinline{reg} statement
in an entity that was described previously.
In a pipeline, the \spadeinline{reg}-statement tells the compiler to insert a pipeline stage, registering
all the values visible above it.
After the \spadeinline{reg}-statement on line 7, any references to \spadeinline{p} will refer to a delayed version
of the value.
The same is true for the \spadeinline{x} value computed by \spadeinline{subpipe}, however, because it
is computed by a pipeline of depth three, the first three registers are present in the sub-pipe
and will not be inserted in the instantiating pipeline.
The times three operator on the register, \spadeinline{reg * 3},  specifies
that three stages should be inserted without any new computations.
Finally, on line 10, \spadeinline{s} is added to the value of
\spadeinline{a} from the first stage. The \spadeinline{stage} keyword is used
to bypass some pipeline
registers and to refer directly to a signal as it appears in another stage, in
this case the stage named \spadeinline{initial}.
Stage references can refer to both ``future'' and ``past'' stages, and can do so by
name, as is done here, or as relative offsets, for example \spadeinline{stage(-2)}.
\begin{lstfloat}
	\begin{lstlisting}[style=error, caption=Example of pipeline error message, label=lst:pipeline_error, autogobble]
	@rerrorr@: Use of x before it is ready
	   @b-->b@ src/main.spade:10:19
	   @b|b@
	10 @b|b@ let sum = x + f(a, product);
	   @b|b@           @r^ Is unavailable for another 2 stagesr@
	   @b=b@ Requesting x from stage 1
	   @b=b@ But it will not be available until stage 3
	\end{lstlisting}
    \vspace{-0.7cm}
\end{lstfloat}

The pipelining feature decouples the description of the computation from the description
of the pipeline itself.
In a pipeline without feedback a developer can easily add and remove pipeline stages as needed without altering the output value of the pipeline.
If such changes potentially affect the outcome of other parts of the project, the inclusion
of the depth at the call site ensures that the user is made aware of such potential issues via
compiler error messages.
In pipelines with feedback the structured description of the stages still helps
during design iteration, even if some manual care is required to ensure correct
computation.

\subsection{Types and Pattern Matching}\label{sub:types}

Spade is a statically typed language with type inference.
This means that types can be omitted in most code since the compiler will infer
the appropriate types from context, and report errors if types can not be
inferred.

Like most languages, Spade has primitive types such as integers and booleans,
and compound types like arrays, tuples, and structs.
In addition, the language supports enum types inspired by software languages
like Rust, Haskell, and ML.\@
Unlike their C or VHDL namesake, these enums have data associated with them in
addition to being one of a set of variants.
A common use case for this
construct is the \spadeinline{Option} type which is defined in the standard
library as shown in Listing~\ref{lst:option}.
It is generic over a contained type \spadeinline{T} and takes on one of two values:
\spadeinline{Some} in which case a value, \spadeinline{val}, of type \spadeinline{T} is present, or
\spadeinline{None} in which case no such value is present.
One can view the \spadeinline{Option} type as having a valid/invalid signal bundled with the data it validates.

\begin{figure}
    \begin{minipage}{0.55\linewidth}
        \begin{lstlisting}[language=spade, autogobble=true, label=lst:option, caption=Definition of the Option type.]
            enum Option<T> {
                None,
                Some{ val: T },
            }
        \end{lstlisting}
        \begin{lstlisting}[language=spade, autogobble=true, caption=Example of pattern matching on a tuple of \spadeinline{Option} values., label=lst:match_example]
        let a: Option<T> = ...;
        let b: Option<T> = ...;
        let result = match (a, b) {
            (Some(val), _) => val,
            (_,  Some(val)) => val,
            _ => 0
        }
        \end{lstlisting}
    \end{minipage}
    \hfill
    \begin{minipage}{0.40\linewidth}
        \centering
        \includegraphics{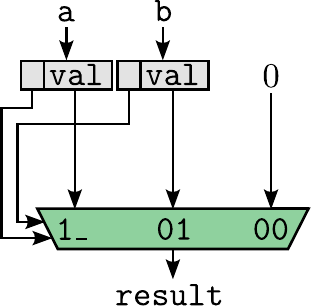}
        \caption{Hardware generated by Listing~\ref{lst:match_example}.}\label{fig:pattern_example}
    \end{minipage}
    \vspace{-0.5cm}
\end{figure}

The main way to interact with enums in Spade is the \spadeinline{match}-expression, which allows pattern
matching on values.
As an example, Listing~\ref{lst:match_example} shows the \spadeinline{match}-expression
in use on a tuple of \spadeinline{Option}-values: \spadeinline{a} and \spadeinline{b}.
The resulting hardware is shown in Fig~\ref{fig:pattern_example}.
If \spadeinline{a} is \spadeinline{Some}, its inner value is bound to \spadeinline{val} and returned from the match expression.
If \spadeinline{a} is \spadeinline{None} but \spadeinline{b} is
\spadeinline{Some}, its inner value is returned.
Finally, if both \spadeinline{a} and \spadeinline{b} are \spadeinline{None}, 0 is returned.
Enums are encoded as bits that specify the currently active variant, the discriminant,
followed by bits containing the payload.
In this case, since the enum consists of two variants, the discriminant is a
single bit while the remaining bits contain the \spadeinline{val} when present and are undefined otherwise.

\subsection{Memories}\label{sub:memories}

As discussed earlier, Spade requires register definitions to
include an expression for the new value of the whole register  as a
function of the values from the previous clock cycle.
However, this abstraction becomes problematic when working with memory-like structures in which
only a small part of the total state is updated in each cycle.
In order to mitigate this, Spade has an explicit construct for memories, currently implemented
as entities defined in the standard library which the compiler handles separately.
Instantiation of a memory is done using the \spadeinline{clocked_memory} entity which creates
a memory with a fixed set of write ports.
Reading from said memory is done via the similarly defined \spadeinline{read_memory} entity.
This means that unlike VHDL and Verilog, memories are explicitly
instantiated as memories, rather than inferred from the structure of the code. An example of using memories is included in the next Section.

\section{Ports}\label{sec:ports}

The discussion so far has been centered around computations on values.
A unit receives a set of values as inputs, and produces a set of values as
output.
Internally, values are used in computation and are pipelined by pipelines.
This becomes inconvenient when working with something like a memory which is
external to the current unit.
The unit must return an address, a write-enable signal, and a value to write
to the memory.
Then the value read from the memory must be passed as an input to the unit.
The memory in turn, produces an output value which must be fed back into the controlling unit.
This approach has a few issues: the control signals and output are delayed by
the pipeline mechanism.
This pipelining introduces additional delays unless manually
mitigated, for example by stage references.
Additionally, there is no clear link between memory control-signals and the
corresponding output.

In order to mitigate this issue, the Spade type system contains the concept of
ports and wires.
Wires come in two forms: mutable and immutable denoted by \spadeinline{&mut} and
\spadeinline{&} respectively.
An immutable wire can be used to pass values via units without them being
delayed in pipelines.
Mutable wires are similarly not pipelined but allow setting the value of the
wire in a module which takes the wire as an input.
Finally, ports are collections of wires and other ports which are bundled together.

Listing~\ref{lst:memory_port} contains an example of how the \spadeinline{port} feature can
be used to share a memory between two pipelines. The first lines define
\spadeinline{port}-types containing read, write, and address wires. Line 8 defines an \spadeinline{entity} where the actual memory is
instantiated. It returns a read-port, and a write-port. Line~9 has been trimmed for space but instantiates the \spadeinline{mut}-wire \spadeinline{r_addr} and a \spadeinline{WPort}.
Lines 10--11 instantiate a memory with a single write-port using the
\spadeinline{clocked_memory} entity, and lines~12--14 asynchronously reads one value from the
memory.
At the end of
the \spadeinline{entity}, the read-port is assembled from the mutable address, and the result
wire, and returned along with the write-port.

The pipelines \spadeinline{reader} and \spadeinline{writer} are two pipelines
which use the read-port and write-port respectively. On line 20, the reader
pipeline sets the address it wants to read from, and reads the resulting value
on line 21. Similarly, the writer sets the target address and write-value on line 27.

Finally, the three units are instantiated on lines 32--34. The ports returned
by the memory are passed to the reader and writer.

\begin{lstfloat}
\begin{lstlisting}[
        language=spade,
        autogobble=true,
        caption=Ports being used to share access to a memory between modules.,
        label=lst:memory_port,
        numbers=left,
        numberstyle=\color{gray},
        xleftmargin=2em
    ]
struct port RPort {
  addr: &mut int<16>, read_val: &int<32>
}
struct port WPort {
  inner: &mut (int<16>, Option<int<32>>)
}

entity dp_mem(clk: clock) -> (RPort, WPort) {
  let (r_addr, w) = ...;
  let w_ports = [inst read_wire(w.inner)];
  let mem = inst clocked_memory(clk, w_ports);
  let r_val = inst read_memory(
      mem, inst read_wire(r_addr)
  );
  (RPort(r_addr, &r_val), w)
}

pipeline(10) reader(clk: clock, r_port: RPort) {
  reg;
    set r_port.addr = address;
    let mem_out = *r_port.read_val;
  reg; // ...
}

pipeline(5) writer(clk: clock, w_port: WPort) {
  reg;
    set w_port.inner = (address, Some(value))
  reg; // ...
}

entity top(clk: clock) -> ... {
  let (r, w) = inst dp_mem(clk);
  let reader_out = inst(10) reader(clk, r);
  let writer_out = inst(5) writer(clk, w);
  // ...
}
\end{lstlisting}
    \vspace{-0.7cm}
\end{lstfloat}

There are some pitfalls when working with mutable wires:
A unit returning a mutable wire expects a value to be set for that wire, otherwise the
value of the wire may be undefined, or if it is set conditionally, a latch could be inferred.
Similarly, if a wire is driven by multiple drivers, conflicting values may cause issues.
While it is possible to catch this in simulation, it is better to let the compiler catch
such errors.

The solution to this is inspired by~\cite{Nigam2020} which uses affine types to
ensure correct memory access patterns in an \ac{HLS} tool. Affine types can
guarantee that a value is used at most once, resolving the multiple driver
problem, however, to ensure that all mutable wires are set exactly once, the
stronger notion of linear types\cite{Wadler1990} is required, and implemented in Spade.

Because Spade describes behavior in a cycle-to-cycle manner, the
implementation of a linear type system is easier than in the general case. Each resource of linear
type  must be \textit{consumed} exactly once each clock cycle.
A value is consumed when it is set using the \spadeinline{set} statement, or passed to
another unit, which delegates the consumption requirement to that unit.

Linear type checking happens after normal type checking, meaning that the compiler
knows which resources are of linear type and must be checked.
For each expression which produces a resource of linear type, a tree is created where leaf
nodes represent primitive linear types, and non-leaves represent compound linear types such as
tuples or structs.
Any statement that aliases a resource, such as a \spadeinline{let}-binding
of an expression, or field access on a struct, creates a pointer from the
alias to the corresponding tree and node.

When expressions or statements which consume resources, such as the \spadeinline{set}-statement,
or a resource being passed to a unit are encountered, the nodes corresponding to the
consumed object are marked as consumed.
If the node or its child nodes are already marked as consumed, the resource
is used more than once and an error is thrown.

This ensures that nodes are not used more than once, but does not guarantee that they are used
exactly once.
Therefore, at the end of the process, a final pass goes over all the trees to ensure that each
node is consumed.
If the traversal finds an unconsumed leaf, it represents a resource of linear
type which was not set, and an error is reported.

\begin{figure}
    \includegraphics{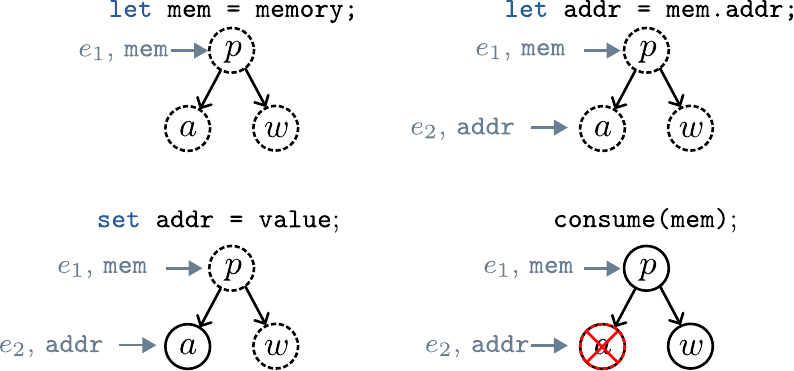}
    \caption{The linear type checking procedure for the code in Listing~\ref{lst:linear_type_checking}.}\label{fig:linear_type_checking}
    \vspace{-0.5cm}
\end{figure}

To exemplify the linear type checking algorithm, Fig.~\ref{fig:linear_type_checking}
shows this process in action for the code in Listing~\ref{lst:linear_type_checking}.
In the figure, $e_x$ represent anonymous names given to sub-expressions before they are
bound to variables. A dashed node represents it not being
consumed, a solid node means it is consumed, and a crossed node
indicates double consumption.

\section{Spade Software}\label{sec:software}
A good set of tools for a language, both for working with the language itself,
and for integration with existing tools can be of huge help for driving language
adoption.
First and foremost, the Spade compiler and the language is built from the start
to produce useful and easy to read error messages such as the one shown in
Listing~\ref{lst:pipeline_error}, and unhelpful diagnostics from the compiler
are considered bugs.
The language is also designed to prefer emitting errors over potentially
surprising behavior, for example, by requiring explicit truncation after
potentially overflowing arithmetic.

\begin{lstfloat}
\begin{lstlisting}[
    language=spade,
    autogobble=true,
    label=lst:linear_type_checking,
    caption=Example code for Fig.~\ref{fig:linear_type_checking}.,
    numbers=left,
	numberstyle=\color{gray},
	xleftmargin=2em
]
    let mem: MemPort = memory();
    let addr = mem.addr;
    set addr = value;
    consume(mem);
\end{lstlisting}
\vspace{-0.7cm}
\end{lstfloat}

\subsection{Compiler Architecture}\label{sec:compiler_architecture}

The Spade compiler is a multi-stage compiler written in Rust, which compiles the
input Spade code to a target language, currently a slim subset of SystemVerilog.
The compilation process starts with lexing and parsing to generate an \ac{AST}.
The \ac{AST} is then traversed thrice, first to collect all types in the
program, then to collect all units, and finally to be lowered into a \ac{HIR}.
The \ac{AST} to \ac{HIR} lowering process retains the tree structure of the \ac{AST} but resolves
names and scoping rules, and performs initial semantic analysis.
Once the \ac{HIR} is generated, type inference is performed, followed by linear
type checking as discussed earlier.
The \ac{HIR} along with the type information is used to generate a \ac{MIR}.
In this step, more semantic analysis is performed, and the tree structure is
flattened to a list of simple statements.
Finally, SystemVerilog is generated from the \ac{MIR}.
SystemVerilog is chosen as the target language as it is well-supported by both open source
and proprietary simulation and synthesis tools, but the compiler is written in a target
independent way to enable experimenting with, or changing to a different
backend with ease.
Especially interesting backend are the CIRCT\cite{CirctWebPage} dialects, such
as LLHD\cite{Schuiki2020} or Calyx\cite{Nigam2021}, which can offer language
independent optimization as well as code generation of other output languages
than SystemVerilog.

\subsection{Tooling and Ecosystem}\label{sec:tooling}

Spade comes with the build tool Swim\footnote{\url{https://gitlab.com/spade-lang/swim}} which
manages project files, backend build tools, and dependencies.
A Swim project consists of Spade files and a configuration file written in
TOML. This file contains, among other things: build tool parameters, raw
Verilog files to be included in the project, and external dependencies.
Swim then manages namespacing of project files, downloading and versioning of
dependencies as well as calling the synthesis and simulation tools with a
convenient interface.
It also has a plugin system for extending the build flow, for example
by running commands to generate Spade or Verilog code, loading additional Yosys
plugins or bundling the output bitstream into an executable for a
microcontroller which in turn programs a target FPGA.

To facilitate integration of Spade with existing projects, units can be
annotated to prevent name mangling. Spade units can also be marked as external,
in order to allow use of existing IP blocks within Spade projects.

A tree-sitter grammar, and a rudimentary \ac{LSP}\cite{lsp_web} server
is available, enabling an IDE-like experience in any text editor
supporting LSP and/or tree-sitter.

There is limited effort to generate Verilog similar in structure to the input
Spade code.
However, the compiler does attempt to keep names readable, and has
functions for mapping names and expressions back to their source location to
aid debugging.
Swim automatically translates values in VCD files into their high-level
Spade values, and the compiler includes a source mapping in the output Verilog
which makes things like timing reports readable without looking at the output Verilog.

\subsection{Test Benches and Simulation}\label{sec:test_benches}

\begin{lstfloat}
	\begin{lstlisting}[
	, language=python
	, autogobble=true
	, caption=Example of a test bench for Spade written using cocotb.
	, label=lst:testbench
	, numbers=left
	, numberstyle=\color{gray}
	, xleftmargin=2em
	]
	# top=peripherals::timer::timer_test_harness
	@cocotb.test()
	async def timer_works(dut):
		s = SpadeExt(dut)

		clk = dut.clk_i
		await start_clock(clk)
		
		s.i.mem_range = "(1024, 2048)"
		s.i.addr = "1024 + 0"
		s.i.memory_command = "Command::Write(10)"
		await FallingEdge(clk)
		s.o.assert_eq("10")
	\end{lstlisting}
    \vspace{-0.6cm}
\end{lstfloat}

The Spade language itself is designed primarily for hardware description and
synthesis, rather than simulation.
However, in order to verify correctness of the resulting hardware, simulation and
test benches are essential.
Spade tests are written using cocotb\cite{cocotb_web}, a Python based
co-simulation test bench environment for verification.
The cocotb library is extended with features for writing Spade values as inputs
and outputs to the unit under test, while Verilog generated by the compiler is simulated with
off-the-shelf Verilog simulators.

As an example, Listing~\ref{lst:testbench} shows part of a test bench for a memory mapped
timer peripheral.
The first line specifies the module being tested, and the next two lines are a
standard cocotb test case definition.
Line~4 wraps the cocotb design under test in a Spade class which extends the cocotb
interface to add Spade-specific features.
Lines~6--7 start a task to drive the clock of the \ac{DUT}.\@
On lines~9--11, the inputs to the module are set to values which are
compiled and evaluated by the Spade compiler.
Finally, line 13 asserts that the output is as expected, again passing a string
containing a Spade expression as the expected value.

In order to achieve this, the Python test bench must be able to compile Spade
code, and in order to allow the use of types and units defined in the project
inside a test bench, the state of the compiler must be made available to the
test bench.
For this reason, the state of the compiler after building a project is serialized and
stored on disk.
Additionally, parts of the Spade compiler are exported as a Python module which
reads the stored state, then compiles and evaluates the expressions used in the test bench.
The resulting bit vectors are used to drive the inputs of the \ac{DUT}, or compared to the expected output.

\section{Conclusions}\label{sec:conclusions}

Spade is an \ac{HDL} which attempts to ease the development of
FPGAs or ASICs. To do so, it makes common hardware constructs like
pipelines, registers, and memories explicit and part of the language.
It is heavily inspired by modern software languages,
e.g., by integrating a powerful type system and pattern matching.
Unlike most current alternative \acp{HDL}, Spade is its own standalone
language with a custom compiler, and does not abstract away the underlying
hardware.
Finally, it comes with useful tooling, such as a compiler designed to emit
detailed error messages, a build system with dependency management, and an \ac{LSP} for an IDE-like experience.

\bibliographystyle{IEEEtran}
\bibliography{IEEEabrv, ../abbreviations/DAabrv, ../main}

\end{document}